\documentclass[a4paper, 12pt]{article}
\author{Clovis Jacinto de Matos\footnote{ESA-HQ, European Space Agency, 8-10 rue Mario Nikis, 75015 Paris, France, e-mail:
Clovis.de.Matos@esa.int} \\
and \\
 Christian Beck\footnote{School of
Mathematical Sciences, Queen Mary, University of London, Mile End
Road, London E1 4NS, UK, e-mail: c.beck@qmul.ac.uk}}
\title{Possible Measurable Effects of Dark Energy in Rotating Superconductors}

\begin{document}

\maketitle \begin{abstract} We discuss recent laboratory
experiments with rotating superconductors and show
that three so far unexplained experimentally observed
effects (anomalous acceleration signals, anomalous gyroscope
signals, Cooper pair mass excess) can be physically explained in
terms of a possible interaction of dark energy with Cooper pairs. Our
approach is based on a Ginzburg-Landau-like model of
electromagnetic dark energy, where gravitationally active photons
obtain mass in the superconductor. We show that this model can
account simultaneously for the anomalous acceleration and
anomalous gravitomagnetic fields around rotating superconductors
measured by Tajmar et al. and for the anomalous Cooper pair mass
in superconductive Niobium, measured by Cabrera and Tate. It is
argued that these three different physical effects are ultimately
different experimental manifestations of the simultaneous
spontaneous breaking of gauge invariance, and of the principle of
general covariance in superconductive materials.
\end{abstract}

\section{Introduction}

The existence of dark energy in the universe, as indicated by numerous astrophysical
observations, represents one of the most challenging problems in theoretical physics
at present
\cite{Spergel,Peebles,Copeland,Padmanabhan}.
A great variety of different models exist for dark energy
but none of these models can be regarded as being entirely convincing
so far. The cosmological constant problem (i.e. the smallness of the
cosmologically observed vacuum energy density)
remains an unsolved problem. It is likely that the solution of this problem
requires new, so far unknown, physics.

While it is clear that dark energy has measurable effects on cosmological scales
(such as the accelerated expansion of the universe as seen from supernovae observations),
it is much less clear what the effects of dark energy could be on smaller scales. These
effects, if any, depend very much on the model considered. For example, if dark energy is due
to the existence of compactified dimensions with a diameter of the order of the micron
scale, then this would lead to modifications of the gravitational interaction
potential on these scales. This can be tested in laboratory
precision experiments. Tests by Adelberger et al. \cite{adelberger} proved negative so far
down to a scale of about 50 microns.

Other more recent models of dark energy, such as the
electromagnetic dark energy model of Beck and Mackey \cite{Beck4},
also produce potentially measurable effects at laboratory scales,
which are, however, restricted to the interior of superconductors.
In \cite{Beck4} a Ginzburg-Landau theory is constructed that generates a
cutoff for the gravitational activity of vacuum fluctuations.
Generally it is assumed in this model that
vacuum fluctuations of any particle can exist in two different phases: A
gravitational active one (contributing to the cosmological
constant $\Lambda$) and a gravitationally inactive one (not
contributing to $\Lambda$).
The model exhibits a phase transition
at a critical frequency which makes the dark energy
density in the universe small and finite. The above approach
%represents a
%low-energy model of dark energy as seen in a fixed reference
%frame (the laboratory). It
has many analogies with the physics of
superconductors, and in particular it allows for a possible
interaction between dark energy and Cooper pairs. A suppression
of the cosmological constant due to the formation of quantum
condensates has also been discussed in \cite{moffat}.

In models of dark energy like the above one it is the {\em new
physics underlying the cutoff} that can potentially lead to
measurable effects in the laboratory (see also \cite{minic}
for related work). In Beck and Mackey's model
\cite{Beck4} dark energy couples to superconducting matter only (and
not to matter in the normal state). This is theoretically
consistent: If we assume that dark energy can interact with
superconducting matter only, we do not get any contradiction from
cosmological observations, since almost all of the matter in the
universe is not in a superconducting state. Given the above
assumption of a possible interaction between dark energy and
superconducting matter one can then constrain the interaction
strength by making precision measurements with superconducting
devices.

In this paper we look at recent experiments that were performed
with rotating superconductors. There are three observed anomalies
that cannot be explained with conventional theories. One dates
back already nearly 20 years: Tate et al. \cite{Tate89} made
precision measurements of the London moment in rotating
superconductors. The London moment is a magnetic field generated
inside a superconductor once it is set into rotation. They
measured a London moment slighty too large as compared to the
theoretical expectations. This anomalous London moment, well
established within the experimental precision, has remained
unexplained for the past 20 years. More recently, Tajmar et al.
\cite{Tajmar2}  investigated rotating superconductors using isolated
laser-gyroscopes positioned outside the superconductor. The
gyroscopes yield small signals proportional to the rotation
frequency that again cannot be explained by conventional theories.
These signals may, however, be interpreted in terms of a
gravitomagnetic field whose strength is much larger than
theoretically expected from ordinary gravity. Finally, Tajmar et
al. \cite{Tajmar1} also measured anomalous induced acceleration signals
in isolated accelerometers close to a rotating superconductor,
which occur if the rotation frequency of the superconductor is
rapidly changed. Again this induced acceleration signal is much
stronger than theoretically expected from normal gravity. All
three effects are specific to superconducting matter only, they
vanish as soon as the temperature exceeds the critical
temperature. Presently it seems that none of the above measured
effects can be understood in terms of conventional superconductor
physics.

We will show that all three effects can be
quantitatively understood from a possible interaction between
Cooper pairs and dark energy as described in the model of Beck and
Mackey \cite{Beck4}. The key for a quantitative understanding of the
experimental observation is the fact that photons formally
obtain mass in a superconductor, due to a finite London
penetration depth. In particular, gravitationally active photons
whose vacuum fluctuations underly dark energy in the model
\cite{Beck4} obtain a mass as well and lead effectively to a strong
enhancement of gravitoelectromagnetic effects, which can be
experimentally measured. Ultimately, our theoretical
interpretation is that not only gauge invariance but also general
covariance is spontaneously broken in the superconducting
material, the latter one being related to an interaction of dark
energy with Cooper pairs.

This paper is organized as follows: In sections 2-3 we briefly
summarize the experimentally observed effects in rotating
superconductors that could possibly be linked to dark energy. In
sections 4-6 we develop the tools for our theoretical approach,
based on the Einstein-Maxwell-Proca equations describing
gravitomagnetic fields and the electromagnetic dark energy model
of Beck and Mackey. Finally, in sections 7-9 we show that our
theory can explain the experimentally observed effects in a
quantitatively correct way.

\section{Tajmar's experiments}
Tajmar et al. have established a research programme at Austrian
Research Centers GmbH -ARC with the objective of peering into new
possible gravitational properties of superconductive materials.
Two different types of experiments have been carried out:

\begin{enumerate}
\item  \label{type_1}The first category provides measurements
of azimuthal accelerations with accelerometers located inside the
equatorial plane of the central hole of different types of
superconductive rings, which are angularly accelerated. The
experimental arrangement was designed to ensure a minimal thermal
and mechanical coupling between the accelerometers and the ring's
motion. The accelerometers were isolated as much as possible from
the whole experimental setup inside a sealed vacuum chamber
\cite{Tajmar2,Tajmar1}.

\item \label{type_2} In the second category, angular velocities were measured with laser gyroscopes
located above different types of uniformly rotating
superconductive rings (with respective equatorial planes parallel
to each other). Like in point \ref{type_1}, the overall
experimental setup and the measurement methods ensured that the
laser-gyros were maximally decoupled from undesired mechanical
torques in general, and from the ring's motion in particular
\cite{Tajmar2}.
\end{enumerate}

Based on the known laws of physics, and taking into account all
known physical effects in the experimental setup described above
in (\ref{type_1}) and (\ref{type_2}), neither the accelerometers
nor the laser-gyros should indicate any significant signal above
the noise level. This is not what Tajmar's experiments
demonstrated. Rather, a clear azimuthal acceleration, which could
be associated with an anomalous gravitational field, directly
proportional to the superconductive ring angular acceleration,
and an angular velocity orthogonal to the ring's equatorial
plane, which could be associated with an anomalous
gravitomagnetic field, have been measured in type-(\ref{type_1})
and type-(\ref{type_2}) experiments respectively
\cite{Tajmar2,Tajmar1}.

\subsection{Acceleration fields around angularly accelerated rotating superconductors}
In Tajmar's experiment, the superconducting ring has a radius of
$R=0.07m$ and the accelerometers are positioned inside the
central hole of the ring at distance $r=0.036m$. In the case of
Niobium rings, the measured coupling between the applied angular
acceleration $\dot{\omega}$ to the superconducting ring (measured
in $rad/s^2$) and the measured induced azimuthal acceleration $g$
inside the central hole of the ring has the following value and
associated error (see table II in \cite{Tajmar1}):
\begin{equation}
\frac{g}{\dot \omega}=-(9.46 \pm 0.28)\times 10 ^{-7}
[m.Rad^{-1}]\label{e1}
\end{equation}
If the angular frequency is measured in units of $[s^{-1}]$, this
result is equivalent to
\begin{equation}
\frac{g}{\dot{\omega}}=- (1.51\pm0.04) \times 10^{-7} [m].
\label{madeleine}
\end{equation}

Like the gravitational field produced by the mass of a physical
body, the measured acceleration, $g$, does not depend on the
accelerometer's mass or chemical composition detecting it.
The effect is only seen if the ring
is in the superconductive state, i.e.\ below its respective
critical temperature $T_c$, and if the ring is accelerated. The
coupling as given by Eq.(\ref{madeleine}) depends on the ring's
material type, and disappears, within the instrumentation
resolution capability, for High-$T_c$ superconductors.

The above value of the coupling is based on single-sensor
measurements and the evaluation of maximum acceleration peaks
\cite{Tajmar1}. If multiple sensors at different
positions are used (in a so-called curl configuration) and if an
average signal analysis is performed, then smaller coupling
constants are obtained \cite{Tajmar2}. However, for the physical
interpretation we are going to provide in section 7 (in terms
of a particle emission event) the maximum
peak analysis with a single sensor
is the most appropriate one.

Acceleration peaks were also observed when the superconductor
passed through its critical temperature while it was rotating at constant
angular velocity. These signals had opposite signs for the transition from
the normal to the superconductive state and vice versa.

%From a general point of view, in this type of experiments, through
%the measurement of accelerations inside the ring's central region,
%we are indirectly capable to assess the effect of the internal
%forces present in the physical system under consideration, i.e.,
%inside the angularly accelerated superconductive ring.

What could be the origin of the measured anomalous acceleration
inside the central region of an angularly accelerated Niobium
superconductive ring? What could account for the measured
coupling, Eq.(\ref{madeleine})?  Before we answer these questions
let us first investigate the anomalous gravitomagnetic properties
of uniformly rotating superconductors.

\subsection{Gravitomagnetic fields around uniformly rotating superconductors}

In the limit of small field strengths and for nonrelativistic
movements, the Einstein equations yield a set of Maxwell-like
equations which describe the so-called gravitomagnetic fields
\cite{Kiefer}. For normal matter these gravitational fields are much
weaker than electromagnetic fields. However, in coherent quantum
systems the experiments seem to provide evidence for much
stronger than expected gravitomagnetic fields \cite{de Matos}.

The most recent experiments of Tajmar et al. use laser-gyroscopes to detect
gravitomagnetic fields.
A laser-gyroscope is an interferometer, measuring the phase
difference between two beams of coherent electromagnetic waves
with equal frequency, $\nu_0$, propagating in opposite directions
along a closed optical fiber. When the fiber rotates
with angular velocity $\Omega$, a phase
difference, $\delta \phi$, is measured. This is the
so-called Sagnac effect:
\begin{equation}
\delta\phi=4 \nu_0 \frac {S\Omega}{c^2}\label{e2}
\end{equation}
Here $S$ is the component of the optical fiber cross section
parallel to the rotation plane.

The same effect, a phase difference, can be caused by a gravitomagnetic flux crossing a
laser gyroscope at rest, since a
gravitomagnetic field $\vec{B}_g$ originates from a vector
potential $\vec{A}_g$:
\begin{equation}
\vec{B_g}=\nabla \times \vec{A_g}\label{e3}
\end{equation}
In that case the phase difference is directly proportional to the
intensity of the gravitomagnetic field:
\begin{equation}
\delta\phi=4 \nu_0 \frac {S B_g}{c^2}\label{e4}
\end{equation}
Here $S$ is the component of the optical fiber cross section
orthogonal to the gravitomagnetic field. Such phase difference are
measured by Tajmar et al. \cite{Tajmar2} when a Niobium
superconductive ring is rotating with constant angular velocity
$\omega$, with an isolated laser-gyroscope positioned nearby.
%The observed phase difference disappears above
%the critical temperature, $T_c$, when the material ceases to be
%superconductive.
The measured coupling $\chi'$ between the
gravitomagnetic field and the angular velocity of the
superconductive Niobium ring is of the order
\begin{equation}
\chi'=\frac{B_g}{\omega}\sim 10^{-8}. \label{e5}
\end{equation}
These gyroscope measurements are not yet finalised and conclusions are still
premature. More precise data are expected to become available
soon \cite{private}.
%It is important to note, for the rest of the discussion, that in
%this type of experiments we are ultimately investigating the
%conservation of the canonical momentum inside the superconductive
%ring.

\section{Cabrera and Tate's measurements}
Superconductors at rest expell any magnetic field. But when they
exhibit rotational motion, a magnetic field $\vec{B}$ is generated
within the superconductor, the so-called London moment:
\begin{equation}
\vec B=-2\frac{m}{e}\vec \omega \label{be11}
\end{equation}
Here $m$ and $e$ are the mass and charge of the Cooper pair. This
effect is not accounted for by classical electrodynamics, it can
only be properly explained in the framework of quantum field
theory\cite{Weinberg1}. It consists in the spontaneous generation
of a magnetic field by setting a superconductor into rotation in
an environment initially (before the rotation starts) entirely
free from any electromagnetic fields.

By measuring the $\vec{B}$ field very precisely, one can conclude
on the Cooper pair mass $m$ using Eq.(\ref{be11}). The
experimental technique for this is based on magnetic flux
quantization: In 1989 Cabrera and Tate \cite{Tate89}, through the
measurement of the London moment magnetic trapped flux, reported
an anomalous Cooper pair mass excess in thin rotating Niobium
superconductive rings:
\begin{equation}
\Delta m=m^*-m=94.147240(21)eV\label{e12}
\end{equation}
Here $m^*=1.000084(21)\times 2m_e=1.023426(21)MeV$ is the
experimentally measured Cooper pair mass (with an accuracy of 21
ppm), and $m=0.999992\times2m_e=1.002331 MeV$ is the theoretically
expected Cooper pair mass including relativistic corrections.
The above Cooper pair mass excess (or, equivalently, the slightly larger than
expected measured magnetic field) has not been explained until now.

Motivated by the absence of any apparent solution of this
disagreement in the existing literature, one of us (CDM)
formulated the conjecture that an additional gravitomagnetic term
must be added to the Cooper pairs' canonical momentum:
\begin{equation}
\vec{\pi}=m\vec v + e \vec A + m \vec {A_g} \label{e13}
\end{equation}
The gravitomagnetic field strength $B_g=|\vec{B}_g|=|\nabla \times
\vec{A}_g|$ required to account for the reported mass excess is 31
orders of magnitude larger than any gravitomagnetic field
predicted by general relativity, based on the mass currents in
the rotating ring\cite{Tajmar03,Tajmar05}. It can be expressed in
terms of Eq.(\ref{e12}) as
\begin{equation}
B_g=\frac{\Delta m}{m}2 \omega=1.84\times 10^{-4} \omega \label{2}
\end{equation}
Here
$\vec{\omega}$ is the superconductor's angular
velocity.
%Eq.~(\ref{2}) is sometimes referred to as the
%"Tajmar-de Matos conjecture".
The physical interpretation of Eq.(\ref{2}) is ambiguous: Either
it can be understood as the gravitomagnetic London-type moment in
rotating superconductors, or it can be associated with a deviation
from the gravitomagnetic Larmor theorem \cite{Mashhoon}, $B_g=2
\omega$, which would be revealed by anomalous Coriolis forces
inside rotating superconductive cavities. These ambiguities will
be further discussed and elucidated below. The most important
question at this point is why the field $B_g$ is so much larger
than expected, and why it is only observed in the superconducting
state. In the following we will relate this to an effect produced
by dark energy.

\section{Massive photons in superconductors}
The properties of superconductors (zero resistivity, Meissner
effect, London moment, flux quantization, Josephson effect, ...)
can be understood from the spontaneous breaking of electromagnetic
gauge invariance when the material is in the superconductive
phase\cite{Weinberg1, Ryder}. In quantum field theory, this
symmetry breaking leads to massive photons via the Higgs
mechanism. In this case the Maxwell equations transform to the
so-called Maxwell-Proca equations, which are given by
\begin{eqnarray}
\nabla \vec{E} =\frac{\rho_e}{\epsilon_0}-\frac{1}{\lambda_\gamma
^2}\phi \label{equ6} \\
\nabla \vec{B}=0 \label{equ7}\\
\nabla\times \vec{E}=-\dot{\vec{B}}\label{equ8}\\
\nabla\times\vec{B}=\mu_0 \rho_e \vec{v}+\frac{1}{c^2}\dot{\vec
E}-\frac{1}{\lambda_\gamma^2}\vec{A}. \label{equ9}
\end{eqnarray}
Here $\vec{E}$ is the electric field, $\vec{B}$ is the magnetic
field, $\epsilon_{0}$ is the vacuum electric permittivity,
$\mu_{0}=1 /\epsilon_0  c^2$ is the vacuum magnetic permeability,
$\phi$ is a scalar electric potential, $\vec{A}$ is a magnetic
vector potential, $\rho_e$ is the Cooper pair condensate charge
density, $\vec{v}$ is the Cooper pair velocity, and
$\lambda_\gamma=\hbar/m_\gamma c$ is the photon's Compton
wavelength, which is equal to the London penetration depth
$\lambda_L=\sqrt{\frac{m}{\mu_o e \rho_e}}$.

Taking the curl of Eq. (\ref{equ9}) and neglecting the term coming
from the displacement current, we get the following equation for
the magnetic field:
\begin{equation}
\nabla^2\vec{B}+\frac{1}{\lambda_\gamma^2}\vec{B}=\frac{1}{\lambda_L^2}\frac{m}{e}2\vec{\omega}.\label{equ10}
\end{equation}
Solving Eq. (\ref{equ10}) for the 1-dimensional case, we obtain a
magnetic field with a term that decays exponentially and another
one that is proportional to the rotation frequency $\omega$. These
are respectively the Meissner effect and the London moment:
\begin{equation}
B=B_0 e^{-x/{\lambda_\gamma}}+2\omega
\frac{m}{e}\Big(\frac{\lambda_\gamma}{\lambda_L}\Big)^2\label{equ11}
\end{equation}
Following Becker et al. \cite{Becker} and London \cite{London},
the London moment is developed by a net current that is lagging
behind the positively charged ion lattice. The Cooper pair
current density shows in opposite direction than the angular
velocity of the superconducting bulk. This is important as the
London moment in all measurements, due to the negative charge of
the Cooper pair, shows in the same direction as the angular
velocity. Having
\begin{equation}
\lambda_{\gamma}=\lambda_L\label{equat12}
\end{equation}
we finally obtain the familiar expression
\begin{equation}
B=B_0 e^{-x/{\lambda_L}}-2\omega \frac{m}{e}.\label{equ13-b}
\end{equation}
The experiments of Tajmar et al. have been operating at
temperatures of the order of $7K$. At this temperature, the
London penetration depth for Niobium is
 $\lambda_L=\lambda_L(0)/\sqrt{1-(T/T_c)^4} =47.6nm$
 (assuming a critical temperature $T_c=9.25K$, and a London
 penetration depth at $T=0$ of $\lambda_{L}(0)=39 nm$
\cite{Kittel}). Substituting this value into Eq. (\ref{equat12}),
we deduce a typical value of the photon mass in Niobium
\begin{equation}
m_\gamma=4.2 eV.\label{equu13}
\end{equation}

\section{Gravitational Maxwell-Proca equations}

In analogy with the electromagnetic fields produced by a Cooper
pair condensate, which are described by the set of Maxwell-Proca
equations Eq.(\ref{equ6})-Eq.(\ref{equ9}), we may write down
analogous equations for gravity in the weak field approximation.
These generate gravitoelectromagnetic fields according to a set of
Einstein-Maxwell-Proca equations, with a massive graviton
\cite{Argyris}:
\begin{eqnarray}
\nabla \vec{g} =-\frac{\rho^*_m}{\epsilon_{0g}}-\frac{1}{\lambda_g
^2}\phi_g \label{equ12} \\
\nabla \vec{B_g}=0 \label{equ13}\\
\nabla\times \vec{g}=-\dot{\vec{B_g}}\label{equ14}\\
\nabla\times\vec{B_g}=-\mu_{0g} \rho^*_m
\vec{v}+\frac{1}{c^2}\dot{\vec g}-\frac{1}{\lambda_g^2}\vec{A_g}
\label{equ15}
\end{eqnarray}
Here $\vec{g}$ is the gravitational field, $\vec{B_g}$ is the
gravitomagnetic field, $\epsilon_{0g}=1/4 \pi G$ is the vacuum
gravitational permittivity, $\mu_{0g}=4\pi G / c^2$ is the vacuum
gravitomagnetic permeability, $\phi_g$ is the scalar gravitational
potential, $\vec{A_g}$ is the gravitomagnetic vector potential,
$\rho^*_m=\rho^*/c^2$ is the mass density of the gravitational
condensate ($\rho^*$ is the corresponding energy density),
$\vec{v}$ is the velocity of the gravitational analogue of Cooper
pairs, and $\lambda_g=\hbar/m_g c$ is the Compton wavelength of
the graviton. At the moment, we will leave open the physical
interpretation of the gravitational quantum condensate, we will
come back to this in the next section. There are several
theoretical approaches for the inclusion of a massive graviton in
general relativity. Here we adopt the proposal from Deser and
Waldron \cite{Deser}, to link the cosmological constant $\Lambda$
and the graviton mass, $m_g$, in a partially massless spin 2
theory in a de-Sitter background ($\Lambda>0)$, which describes 4
propagating degrees of freedom for the graviton corresponding to
helicities $(\pm2,\pm1)$ (helicity 0 being unphysical). In this
approach one has
%Novello and others \cite{Novello1,
%Novello2,Liao} to link the cosmological constant
%$\Lambda$ and the graviton mass, which arises as a natural
%consequence of the equations of motion of a massive graviton
%propagating in a de-Sitter background. In this approach one has
\begin{equation}
\frac{1}{\lambda_g^2}=\Big(\frac{m_g
c}{\hbar}\Big)^2=\frac{2}{3}\Lambda .\label{5c}
\end{equation}
%(strictly speaking, there is a minus sign in this equation,
%meaning one has an imaginary graviton mass).
At this particular (critical) value a novel local scalar gauge
invariance appears that is responsible for the elimination of the
helicity 0 excitation. In this theory in the limit of a massless
field, $m_g=0$, the classical two degrees of freedom $\pm2$ are
recovered, and in the interval
$0<m^2_g<\frac{2\hbar^2\Lambda}{3c^2}$ the theory is unstable.
Note that the above graviton mass is tiny: $\lambda_g/c$ is of the
same order of magnitude as the current age of the universe, hence
we should not expect any significant modification with respect to
the presently experimentally verified laws of gravity from such a
tiny postulated graviton mass.

In the 1-dimensional case we obtain the solution of
Eq.~(\ref{equ15}) as
\begin{equation}
B_g={B_g}_0e^{-x/\lambda_g} +2 \omega \left( \frac{\lambda_g}{\lambda_L} \right)^2
\label{gravi}
\end{equation}
where
\begin{equation}
\lambda_L=\frac{1}{\sqrt{{\mu_0}_g\rho_m^*}}
\end{equation}
is the gravitational analogue of the London penetration depth. In
Eq.~(\ref{gravi}) the $x$-dependence of the first term (the
gravitational Meissner effect) can be neglected due to the huge
$\lambda_g$ assumed. The term of interest is the second term, the
gravitomagnetic London-type moment, which can potentially lead to
measurable effects, depending on what is assumed for $\rho^*$. The
basic idea in the following is to associate $\rho^*$ with dark
energy.

\section{Dark energy in a superconductor}

So far the Einstein-Maxwell Proca equations considered in the
previous section were not coupled to the electrodynamics of Cooper
pairs. We now consider possible interactions,
based on the model of dark energy as proposed by Beck and Mackey in \cite{Beck4}.

A non-vanishing cosmological constant (CC) $\Lambda$ can be interpreted in
terms of a non-vanishing vacuum energy density
\begin{equation}
\rho_{vac}=\frac{c^4}{8\pi G} \Lambda ,\label{e14}
\end{equation}
which corresponds to dark energy with equation of state $w=-1$.
The small astronomically observed value of the CC,
$\Lambda=1.29\times10^{-52}[1/m^2]$ \cite{Spergel}, and its origin
remain a deep mystery. This is often call the CC problem, since
with a cutoff at the Planck scale the vacuum energy density
expected from quantum field theories should be larger by a factor
of the order $10^{120}$, in complete contradiction with the
observed value. To solve the CC problem, in \cite{Beck4} a model
of dark energy was suggested that is based on electromagnetic
vacuum fluctuations creating a small amount of vacuum energy
density. One assumes that photons (or any other bosons), with
zeropoint energy $\epsilon=\frac{1}{2} h \nu$, can exist in two
different phases: A gravitationally active phase where the
zeropoint fluctuations contribute to the cosmological constant
$\Lambda$, and a gravitationally inactive phase where they do not
contribute to $\Lambda$ \cite{Beck4, Beck,Beck2,Beck3}. This is
described in \cite{Beck4} by a Ginzburg-Landau type of theory. As
shown in \cite{Beck4}, this type of model of dark energy can lead
to measurable effects in superconductors, via a possible
interaction with the Cooper pairs in the superconductor.

Here we introduce the following additional
hypotheses with respect to the original Beck and Mackey model, which, as
we shall see later, are consistent with the experimental observations
in rotating superconductors:

\begin{enumerate}
\item\label{i1} Like normal photons (with energy
$\epsilon=h\nu$), the gravitationally active photons (with
zeropoint energy $\epsilon=\frac{1}{2} h \nu$) acquire mass in a
superconductor due to the spontaneous breaking of gauge
invariance. In the following we call these spin 1 bosons
\emph{graviphotons}. In Niobium their mass is approximately $\sim
4 eV$, as we saw previously in Eq.(\ref{equu13}).

\item \label{i2}The transition between the two
graviphoton's phases (gravitationally active, versus
gravitationally inactive) occurs at the critical temperature
$T_c$ of the superconductor, which defines a cutoff frequency $\nu_c$ of
zeropoint fluctuations specific
to each superconductive material: $h\nu_c \sim kT_c$.

\item \label{i3} Graviphotons can form weakly bounded states with Cooper
pairs, increasing their mass slightly from $m$ to $\tilde{m}$. The binding
energy is $\epsilon_c=\mu c^2$:
\begin{equation}
\tilde{m}=m+m_{\gamma}-\mu\label{e155}
\end{equation}

\item \label{i4} Since the graviphotons are bounded to the Cooper pairs, their
zeropoint energies form
a condensate capable of the gravitoelectrodynamic properties of
superconductive cavities.

\end{enumerate}

Beck and Mackey's Ginzburg-Landau-like theory leads to a finite
dark energy density dependent on the frequency cutoff $\nu_c$ of
vacuum fluctuations:
\begin{equation}
\rho^*=\frac{1}{2}\frac{\pi h}{c^3}\nu_c^4\label{e15}
\end{equation}
In vacuum one may put
$\rho^*=\rho_{vac}$, from which the cosmological cutoff frequency
$\nu_{cc}$ is estimated as
\begin{equation}
\nu_{cc}\simeq2.01 THz\label{e16}
\end{equation}
The corresponding "cosmological" quantum of energy is:
\begin{equation}
\epsilon_{cc}=h\nu_{cc}=8.32 meV\label{e166}
\end{equation}
In the interior of superconductors, according to assumption 2.,
the effective cutoff frequency can be different. This is due to
interaction effects between the two Ginzburg-Landau potentials
(that of the superconducting electrons and that of the dark energy
model) \cite{Beck4}. The effect can be seen in analogy to
polarization effects of ordinary electromagnetic fields in matter:
%In our model the superconductor has the ability to `polarize' dark energy.
In matter the electric field energy density is different as compared to the vacuum.
Similarly, in superconductors the effectice dark energy density
(represented by gravitationally active zeropoint fluctuations)
can be different as compared to the vacuum. Our model allows for
the gravitational analogue of polarization.

An experimental effort is currently taking place at University College London and
the University of Cambridge to measure the cosmological cutoff frequency through
the measurement of the spectral density of the noise current in
resistively shunted Josephson junctions, extending earlier
measurements of Koch et al. \cite{Koch}.

In \cite{Beck4}
the formal attribution of a temperature $T$ to the graviphotons is
done by comparing their zeropoint energy with the energy of ordinary photons in a
bath at temperature $T$:
\begin{equation}
\frac{1}{2} h\nu=\frac{h\nu}{e^{\frac{h\nu}{kT}}-1}\label{e16a}
\end{equation}
This condition is equivalent to
\begin{equation}
h\nu=\ln3kT\label{e17}
\end{equation}
Substituting the critical transition temperature $T_c$ specific to
a given superconductive material into Eq.(\ref{e17}), we can
calculate the critical frequency characteristic for this material:
\begin{equation}
\nu_c=\ln3 \frac{kT_c}{h}\label{e18}
\end{equation}
For example, for Niobium with $T_c=9.25$K we get $\nu_c=0.212$
THz. If we use the cosmological cutoff frequency in Eq.(\ref{e18})
we find the cosmological critical temperature $T_{cc}$:
\begin{equation}
T_{cc}=87.49K\label{e19}
\end{equation}
This temperature is characteristic of the BSCCO High-$T_c$
superconductor.

\section{Graviphotonic effect in accelerated superconductors}

Let us now come to our physical explanation of the observed
experimental effects, using the dark energy model of the previous
section. In Tajmar's type-1 experiments, a strong angular
acceleration applied to the superconductive ring can break the
bound between a Cooper pair and its associated graviphoton. In
that process the Cooper pair looses the mass $m_{\gamma}\sim 4eV$,
which by reaction effectively produces an acceleration, on the
Cooper pairs, in the opposite direction of the applied
acceleration. This generates a macroscopic, measurable
acceleration, since a coherent patch of a large number of
graviphotons are simultaneously expelled out of the superconductor
and absorbed by the vacuum. Our hypothesis is that this produces
the measured induced acceleration in Tajmar's experiment, as can
be deduced by applying Newton's law of action-reaction to the
system formed by a Cooper pair and the graviphoton:
\begin{equation}
-f_{Graviphoton}=-m_{\gamma} \dot{\omega} R=mg=f_{Cooper Pair}
\label{1e2}
\end{equation}
Here $R\sim0.07 [m]$ is the superconductive Niobium ring radius
(in Tajmar's experiment), $f_{Graviphoton}$ is the force applied
on the massive graviphotons due to the angular acceleration
communicated to the superconductive ring, and $f_{Cooper Pair}$ is
the corresponding reaction force experienced by the Cooper pair,
$\dot \omega$ is the superconductor's angular acceleration
measured in $s^{-2}$, and $g$ is the produced acceleration at
distance $R$. From Eq.(\ref{1e2}), we find the coupling
\begin{equation}
\frac{g}{\dot{\omega}}=-\frac{m_{\gamma}}{m} R
.\label{1e3}
\end{equation}
Substituting into Eq.(\ref{1e3}) the value of the graviphoton mass
for Niobium found in Eq.(\ref{equu13}), we obtain at distance
$R=0.07m$ the value ${g}/{\dot{\omega}} \approx -2.9 \times
10^{-7} [m]. $ But the actual acceleration measurements are done
with accelerometers that are positioned at the smaller distance
$r=0.036m$, where according to Eq.~(\ref{equ14}) we expect to see
an induced field that is smaller by a factor $r/R=0.51$. We thus
obtain at $r=0.036m$ the theoretical prediction
\begin{equation}
\frac{g}{\dot{\omega}} \approx -1.49 \times 10^{-7} [m].
\end{equation}
This is in
 excellent agreement with the experimentally measured
value as given in Eq.~(\ref{madeleine}), $g/ \dot{\omega}=-
(1.51\pm 0.04) \times 10^{-7} [m]$.

If the superconductive ring rotates with constant angular
velocity and the temperature $T$ is decreased below $T_c$, then
spontaneously Cooper pairs form and these immediately absorb
graviphotons out of the vacuum. By this process their mass
increases, and an acceleration signal is detected. Similarly, if
$T$ is increased above $T_c$, then the material immediately
looses the graviphotons to the vacuum, the Cooper pair mass
decreases, and an opposite acceleration signal is measured, as
reported in Tajmar's experiment.

We suggest to call this coherent emission of graviphotons by accelerated
superconductors the "\emph{graviphotonic effect}".
%Another possible name would be  "\emph{gravitomagnetic
%Bremsstrahlung}".

\section{Gravitomagnetic London moment in rotating superconductors}

Let us first provide a short derivation
of the ordinary electromagnetic London moment---after that we will
proceed to the gravitomagnetic London moment.
The Cooper pairs in a superconductor can be regarded
as a condensate described by a single wave function $\Psi$:
\begin{equation}
\Psi=\rho_e^{1/2}e^{i\phi}\label{e6}
\end{equation}
Here $\rho_e$ is equal to the electric charge density of the
Cooper pair condensate, and $\phi$ is the phase of the wave
function. The canonical momentum of the Cooper pairs, $\vec{\pi}$,
is proportional to the gradient of the phase of the wave function
\begin{equation}
\vec{\pi} \sim \nabla\phi .\label{e7}
\end{equation}
Explicitly, the canonical momentum contains a mechanical and a
magnetic term:
\begin{equation}
\vec{\pi}=m\vec{v} + e\vec{A}\label{e8}
\end{equation}
Here again $m \approx 2m_e$ is the Cooper pair mass and $e$ the
Cooper pair charge (i.e.\ twice the electron charge), $\vec{v}$ is
the Cooper pair's velocity with respect to an inertial reference
frame attached to the laboratory, and $\vec{A}$ is the magnetic
vector potential. From Eq.(\ref{e7}) we deduce that the curl of
the Cooper pairs' canonical momentum is always zero:
\begin{equation}
\nabla\times\vec{\pi}=0\label{e9}
\end{equation}
%Substituting Eq.(\ref{e8}) into Eq.(\ref{e9}) we arrive at the
%well known second London equation:
%\begin{equation}
%\vec{v}=-\frac{e}{m}\vec A\label{e10}
%\end{equation}
Taking the curl of Eq.(\ref{e8}) and using the fact that
$\vec{v}=\vec{r} \times \vec{\omega}$ and $\vec{B}=\nabla \times
\vec{A}$  we obtain the London moment
\begin{equation}
\vec B=-2\frac{m}{e}\vec \omega . \label {e11}
\end{equation}

In close analogy to the above derivation, let us now proceed to
a gravitomagnetic London moment as produced by dark energy.
As seen before, our central hypothesis is that the gravitational quantum
condensate is related to dark
energy.
We may assume that cosmological quanta of energy $h\nu_{cc}$
manifest themselves as massive particles of mass $\mu$
in the superconductor:
\begin{equation}
\mu c^2 = h\nu_{cc}
\end{equation}
These can be regarded as the gravitational analogues of the
Cooper pairs in our model (note the similarity with typical
neutrino mass scales). We expect them to be strongly correlated
to ordinary Cooper pairs and to rotate with the superconductor.
In close analogy to eq.~(\ref{e8}) (i.e.\ replacing $m\to \mu$,
$e\to -m$, $\vec{A} \to \vec{A_g}$) we now consider the
gravitational canonical momentum
\begin{equation}
\vec{\pi_g}= \mu \vec{v} -m\vec{A_g}-m_\gamma \vec{A_g},
\label{45}
\end{equation}
where the term $-m \vec{A_g}$ describes the interaction with the
ordinary Cooper pairs of mass $m$. Assuming that the dark energy
quanta, the ordinary Cooper pairs and the graviphotons are
described by the same macroscopic wave function (i.e.
implementing the assumption of phase synchronisation as in
\cite{Beck4}) we again obtain $\vec{\pi_g} \sim \nabla \Phi_g$
where $\Phi_g$ is the gravitational phase, hence
\begin{equation}
\nabla \times \vec{\pi_g}=0.
\end{equation}
By taking the curl in Eq.(\ref{45}) we thus obtain the
gravitomagnetic London moment
\begin{equation}
\vec{B_g}= 2 \frac{\mu}{m+m_\gamma} \vec{\omega}.
\end{equation}
Putting in $m+m_\gamma \approx m \approx 2m_e$ and $\mu c^2
=h\nu_{cc}$ we obtain the numerical value
\begin{equation}
B_g=  1.62 \times 10^{-8} \omega .
\end{equation}
This theoretical prediction is consistent with Tajmar's type 2
measurements, Eq.(\ref{e5}), where ${B_g}/{\omega}$ was measured
to be of the order $10^{-8}$.

%Due to the close analogy between this phenomena and the well known
%London moment, we propose to designate this effect as the
%"\emph{Gravitomagnetic London Moment}"

\section{Non-classical inertia in superconductive cavities}
From the Einstein-Maxwell-Proca equations of our
electromagnetic model of dark energy with massive bosons we can derive the inertial
properties of a superconductive cavity. Taking the gradient of Eq.
(\ref{equ12}), and the curl of Eq. (\ref{equ15}), and
solving the resulting differential equations for the
1-dimensional case we find respectively a form of the principle of
equivalence and of the gravitomagnetic Larmor
Theorem\cite{Mashhoon} in superconductive cavities:
\begin{eqnarray}
\vec g=-\vec a \mu_{0g} \rho^*_m \lambda^2_g\label{equ16}\\
\vec B_g= 2\vec\omega\mu_{0g} \rho^*_m \lambda^2_g \label{equ17}
\end{eqnarray}
Here $\vec{a}$ is an acceleration communicated to the
superconductive cavity, and $\vec{g}$ is an acceleration measured
inside the superconductive cavity. For the derivation of
Eq.~(\ref{equ16}) we assumed the case of a homogeneous field
$\vec{g}$ and used the formulas $\vec{g}=-\nabla \phi_g$ and
$\rho_m^* \vec{a}=-\nabla(\rho_m^*c^2)$. For Eq. (\ref{equ17}) we
used Becker's argument that the Cooper pairs are lagging behind
the lattice so that the current is flowing in the opposite
direction of $\omega$. We now express the coupling
$\chi=B_g/\omega$ between the gravitomagnetic field and the
superconductor's angular velocity directly as a function of the
dark energy density $\rho^*$ contained in the superconductor
through the substitution of $\rho_m^*=\rho^*/c^2$ in
Eq.(\ref{equ17}):
\begin{equation}
\chi=\frac{8\pi G}{c^4}\rho^*\lambda_g^2\label{4}
\end{equation}
Substituting Eq.(\ref{e14}) and Eq.(\ref{5c}) into Eq.(\ref{4})
and rearranging we obtain
\begin{equation}
\chi=\frac{3}{2}\frac{\rho^*}{\rho_{vac}}.\label{7}
\end{equation}
Substituting Eq.(\ref{e14}), Eq.(\ref{e15}) and Eq.(\ref{e18})
into Eq.(\ref{7}) we obtain
\begin{equation}
\chi=\frac{3\ln^4 3}{4 \pi}\frac{k^4G}{c^7\hbar^3 \Lambda}
T_c^4\label{12}.
\end{equation}
Remarkably, this equation connects the five fundamental
constants of nature $k,G,c,\hbar, \Lambda$ with measurable
quantities in a superconductor, $\chi$ and $T_c$.

We may define a Planck-Einstein temperature $T_{PE}$ as
\begin{equation}
T_{PE}=\frac{1}{k}\Bigg(\frac{c^7\hbar^3
\Lambda}{G}\Bigg)^{1/4}=60.71 K. \label{13}
\end{equation}
Eq.(\ref{12}) can then be written as \cite{de Matos4}
\begin{equation}
\chi=\frac{3\ln^4 3}{4
\pi}\Bigg(\frac{T_c}{T_{PE}}\Bigg)^4.\label{14}
\end{equation}
Substituting the critical transition temperature of Niobium,
$T_c=9.25K$, into Eq.(\ref{14}) we find the following
coupling between the gravitomagnetic field and the angular
velocity of a rotating superconductive Niobium ring:
\begin{equation}
\chi=1.87\times10^{-4}\label{15}
\end{equation}
This coupling describes the effect of gravitationally active
zeropoint energy within the superconductor.
The above theoretically predicted value is extremely close to the
measured value in Cabrera and Tate's experiment, Eq.(\ref{2}):
\begin{equation}
\chi=2\frac{\Delta m}{m}=1.84\times10^{-4}\label{16}
\end{equation}

Let us evaluate the theoretically predicted
coupling for various types of superconductors,
starting with Aluminium and ending with High-$T_c$ superconductors like
YBCO:

\begin{center}
\begin{tabular}{|c|c|c|}
\hline Superconductive material & $T_c [K]$ & $\chi$ \\
\hline $Al$ & $1.18$ & $4.96\times10^{-8}$ \\ \hline $In$ & $3.41$ & $3.46\times10^{-6}$ \\
\hline $Sn$ & $3.72$ & $4.90\times10^{-6}$ \\
\hline $Pb$ & $7.2$ & $6.88\times10^{-5}$ \\
\hline $Nb$ & $9.25$ & $1.87\times10^{-4}$ \\
\hline High-$T_c$ & $79.06$ & $1$ \\
\hline $BSCCO$ & $87.5$ & $1.5$ \\
\hline $YBCO$ & $94.0$ & $2$ \\ \hline
\end{tabular}
\end{center}
Table 1: Predicted coupling $\chi$ between the gravitomagnetic field and the
angular velocity for different superconductive materials.
\bigskip

We note that for YBCO, with $T_c=94.0K$, we recover the classical
gravitational Larmor theorem \cite{Mashhoon}:
\begin{equation}
B_g=2\omega\label{16b}
\end{equation}
From Table 1 we conclude that the effective laws of inertia in
superconductive cavities deviate from the laws of classical
mechanics, recovering however the classical regime in the limit
of YBCO cavities.

At this point one remark is at order:
Our theoretical
derivation presented in this paper strictly speaking holds only
for conventional low-$T_c$ superconductors, because we are using simple Ginzburg-Landau
models and BCS-type of arguments for both the superconductor and the dark energy model \cite{Beck4}.
A high-$T_c$ superconductor, however, is not decribed by such a simple
theory. Whether a high-$T_c$ superconductor can form bounded states with
graviphotons, thus leading to our predicted effects, is theoretically unclear.
The measurements of Tajmar do not show any signal above the noise level
for high-$T_c$ superconductors. Moreover, the London moment measured for
high-$T_c$ superconductors seems not to show any anomalous behavior
within the experimental precision \cite{verhei}. It is likely
that our theory holds for conventional superconductors only,
where the coupling $\chi$ is small.

Our approach raises interesting perspectives for future experiments:
\begin{enumerate}
\item Measurements of the Coriolis force in rotating
low-$T_c$ superconductive cavities should show a deviation from
the predictions of classical mechanics.

\item If in Tajmar's type-2 experiment the laser-gyro is mechanically
attached to the rotating Niobium superconductive ring we should
find the "gravito-angular" coupling, $\chi=1.84\times10^{-4}$,
associated with the Cabrera and Tate Cooper pair mass excess. If
instead the laser-gyro is mechanically decoupled from the rotating
superconductive ring, which correspond to the current
configuration, we should observe the gravitomagnetic London moment
characterised by a coupling $\chi'=1.61\times10^{-8}$,
Eq.(\ref{e5}), consistent with Tajmar's recent experiments.

\item The dependence of the coupling $\chi$ on various superconducting materials
should be further inverstigated. High-$T_c$ superconductors may require a different
gravitomagnetic theory.
\end{enumerate}

\section{Discussion}
Let us end this paper with some general remarks.
General Relativity is founded on the \emph{Principle of
Equivalence}, which rests on the equality between the inertial and
the gravitational mass of any physical system.
%, and formulates that
%\emph{at every space-time point in an arbitrary gravitational
%field it is possible to choose a "locally inertial coordinate
%system" such that, within a sufficiently small region of the point
%in question, the laws of nature take the same form as in
%unaccelerated Cartesian coordinate systems in the absence of
%gravity}. In other words, The inertial frames, that is, the
%"freely falling coordinate systems", are indeed determined by the
%local gravitational field, which arises from all the matter in the
%universe, far and near. However, once in an inertial frame, the
%laws of motion are completely unaffected by the presence of nearby
%masses, either gravitationally or in any other way.
The \emph{Principle of General
Covariance} (PGC) is an alternative version of the principle of
equivalence\cite{Weinberg}, which is very appropriate to
investigate the field equations for electromagnetism and
gravitation. It states that \emph{a physical equation holds in a
general gravitational field if two conditions are met}:
\begin{enumerate}
\item The equation holds in the absence of gravitation; that is, it
agrees with the laws of special relativity when the metric tensor
$g_{\alpha\beta}$ equals the Minkowsky tensor $\eta_{\alpha\beta}$
and when the affine connection $\Gamma_{\beta\gamma}^{\alpha}$
vanishes.
\item The equation is generally covariant; that is, it preserves
its form under a general coordinate transformation $x \rightarrow
x'$.
\end{enumerate}

%It should be stressed that general covariance by itself is empty
%of physical content. The significance of the principle of general
%covariance lies in its statement about the effects of gravitation,
%that a physical equation by virtue of its general covariance will
%be true in a gravitational field if it is true in the absence of
%gravitation. The PGC is not an invariance principle, like the
%principle of Galilean or special relativity, but is instead a
%statement about the effects of gravitation, and about nothing
%else. In particular general covariance does not imply Lorentz
%invariance.
Any physical principle such as the PGC, which takes the form of an
invariance principle but whose content is actually limited to a
restriction on the interaction of one particular field, is called
a dynamic symmetry. Local gauge invariance, which governs the
electromagnetic interaction, is an important example of a dynamic
symmetry. We can actually say that the Principle of General
Covariance in general relativity is the analogon of the Principle
of Gauge Invariance in electrodynamics. The breaking of gauge
invariance leads to superconducting states. The breaking of
general covariance leads to non-conservation of energy-momentum
(in the covariant sense)\cite{Weinberg}. In this context it is
interesting to see that our gravitational dark energy quantum
condensate is related to zeropoint fluctuations, for which energy
is certainly not conserved.

\section{Conclusion}

In this paper we have investigated in detail the possibility that the
dark energy of the universe may interact with Cooper pairs in superconductors,
thus leading to effects that can be observed in the laboratory. Whether or
not such an interaction is a realistic assumption depends very much on the
dark energy model considered. The electromagnetic dark energy model of Beck and Mackey
\cite{Beck4}, and its further development as presented
in the current paper, naturally contains such an interaction.

There are first experimental hints that one might be on the right
track with these types of theoretical models. The graviphotonic
effect, the gravitomagnetic London moment, and non-classical
inertia in rotating superconductive cavities are three different
experimentally observed effects which can all be explained by the
proposed model of dark energy --- not only qualitatively but also
quantitatively. The model ultimately relies on the spontaneous
breaking of gauge invariance and the spontaneous breaking of the
principle of general covariance in the interior of
superconductors.

The considerations presented in this paper, if confirmed by
further independent experiments, would imply that the dark energy
of the universe produces measurable effects not only on cosmological
scales but also in the interior and the vicinity of superconductors.
This opens up the way for a variety of new possible laboratory experiments testing
the nature of dark energy and constraining the interaction
strength with Cooper pairs.
In our model
gravitationally active vacuum fluctuations underlying
dark energy lead to a strong enhancement of gravitomagnetic
fields, in quantitative agreement with
the anomalies seen in the
experiments of Tate et al.\cite{Tate89} and Tajmar et
al.\cite{Tajmar2, Tajmar1}.

\section{Acknowledgments}

We would like to thank Dr. Martin Tajmar for useful comments on an
early version of this paper.
 C.B. is supported by a Springboard
Fellowship of EPSRC.


\begin{thebibliography}{99}

\bibitem{Spergel} D. N. Spergel et al., Astrophys. J. Suppl.
\textbf{148}, 175 (2003)

\bibitem{Peebles} P. J. E. Peebles, B. Ratra, {\it Rev. Mod.
Phys.} {\bf 75}, 559 (2003)

\bibitem{Copeland} E.J. Copeland, M. Sami, and S. Tsujikawa,
{\it Int. J. Mod. Phys.} D {\bf 15}, 1753 (2006)

\bibitem{Padmanabhan} T. Padmanabhan, {\it Phys. Rep.} {\bf 380},
235 (2005)

\bibitem{adelberger} D. Kapner et al., {\it Phys. Rev. Lett.} {\bf 98}, 021101 (2007)

\bibitem{Beck4} C. Beck, M. C. Mackey, "Electromagnetic Dark Energy",
astro-ph/0703364

\bibitem{moffat} S. Alexander, M. Mbonye, and J. Moffat,
"The Gravitational Instability of the Vacuum: Insight into the
Cosmological Constant Problem", hep-th/0406202

\bibitem{minic} V. Jejjala, M. Kavic, and D. Minic, "Fine structure of
dark energy and new physics", arXiv:0705.4581

\bibitem{Tate89} J. Tate, B. Cabrera, S. B. Felch, J. T. Anderson,
{\it Phys. Rev. Lett.} {\bf 62}, (8) 845 (1989).

\bibitem{Tajmar2}M. Tajmar, F. Plesescu, B. Seifert, K. Marhold,
"Measurement of gravitomagnetic and acceleration fields around a
rotating superconductor", AIP Conf. Proc. {\bf 880}, 1071 (2007)
[gr-qc/0610015]

\bibitem{Tajmar1}M. Tajmar, F. Plesescu, K. Marhold, C. J. de Matos, "Experimental Detection of the Gravitomagnetic London
Moment", gr-qc/0603033

\bibitem{Kiefer} C. Kiefer, C. Weber, {\it Annalen Phys.} {\bf 14},
253 (2005).

\bibitem{de Matos} C. J. de Matos, M. Tajmar, {\it Physica C} {\bf
432}, 167 (2005).

\bibitem{private} M. Tajmar, work in progress,
to appear in the proceedings of the GRG18 conference (2007)

\bibitem{Weinberg1} S. Weinberg, {\it The Quantum Theory of
Fields}, Vol 2, (Cambridge University Press, 1996), p. 332.

\bibitem{Tajmar03} M. Tajmar, C. J. de Matos, {\it Physica C} {\bf
385}, {551} (2003).

\bibitem{Tajmar05} M. Tajmar, C. J. de Matos, {\it Physica C} {\bf
420}, {56} (2005).

\bibitem{Mashhoon} B. Mashhoon, {\it Phys. Lett. A} \textbf{173}, 347
(1993)

\bibitem{Ryder} L. H. Ryder, {\it Quantum Field Theory}, 2nd edn.
(Cambridge University Press, 1996), p. 296.

\bibitem{Becker}R. Becker, G. Heller, F. Sauter, {\it Z. Physik} {\bf 85},
772 (1933).

\bibitem{London}F. London, {\it Superfluids}, (John Wiley and
Sons, New York, 1950).

\bibitem{Kittel} C. Kittel, {\it Introduction to Solid State Physics}, 8th Ed., (John Wiley and Sons, New York 2005), p. 275.

\bibitem{Argyris}J. Argyris, {\it Aust. J. Phys.} {\bf 50}, 879
(1997).

\bibitem{Deser}S. Deser, A. Waldron,
%"Stability of massive
%cosmological gravitons",
{\it Phys. Lett. B} {\bf 508}, 347 (2001)  [hep-th/0103255v2]

\bibitem{Beck} C. Beck, M. C. Mackey, {\it Phys. Lett. B} {\bf
605}, 295 (2005).

\bibitem{Beck2} C. Beck, {\it J.Phys.Conf.Ser.} {\bf 31}, 123 (2006).

\bibitem{Beck3} C. Beck, M. C. Mackey, {\it Physica A} {\bf 379}, 101
(2007)

\bibitem{Koch} R. H. Koch, D. van Harlingen, and J. Clarke, {\it Phys.
Rev. Lett.} \textbf{45}, 2132 (1980)

\bibitem{de Matos4} C. J. de Matos, "Electromagnetic dark energy
and gravitoelectrodynamics in superconductors", arXiv 0704.2499,
submitted to Phys. Lett. B (2007)

\bibitem{verhei} A.A. Verheijen et al., {\it Nature} {\bf 345}, 418 (1990)

\bibitem{Weinberg} S. Weinberg, {\it Gravitation and Cosmology:
Principles and Applications of the General Theory of Relativity},
(John Wiley and Sons, New York, 1972) p.91, 111, 361.

%%%%%%%%%%%%%%%%%%%%%%%%%%%%%%%%%%%%%%%%%%%%%%%%%%%%%%%%%%%%%%%%%%%%%%%%%%%%%%%%%%%%%%%%
%%%%%%%%%%%%%%%%%%%%%%%%%%%%%%%%%%%%%%%%%%%%%%%%%%%%%%%%%%%%%%%%%%%%%%%%%%%%%%%%%%%%%%%%%
%%%%%%%%%%%%%%%%%%%%%%%%%%%%%%%%%%%%%%%%%%%%%%%%%%%%%%%%%%%%%%%%%%%%%%%%%%%%%%%%%%%%%%

%\bibitem{Tate90} J. Tate, B. Cabrera, S. B. Felch, J. T. Anderson,
%{\it Phys. Rev. B} {\bf 42}, (13) 7885 (1990).



%\bibitem{Tajmartbp} M. Tajmar, to be published

%\bibitem{pg} Particle data group, http://pdg.lbl.gov

%\bibitem{He} X.-G. He, A. Zee, hep-ph/0702133

%\bibitem{Koch} R.H. Koch, D. van Harlingen, and J. Clarke,
%{\it Phys. Rev. Lett. } {\bf 47}, 1216 (1981)

%%%%%%%%%%%%%%%%%%%%%%%%%%%%%%%%%%%%%%%%%%%%%%%%%%%%%%%%%%%%%%%%%%%%%%%%%%%%%

%\bibitem{Liu} M. Liu, {\it Phys. Rev. Lett.} {\bf 81}, (15) 3223
%(1998).

%\bibitem{Jiang} Y. Jiang, M. Liu, {\it Phys. Rev. B} {\bf 63},
%184506 (2001).

%\bibitem{Capellmann} H. Capellmann, {\it Europ. Phys. Jour. B}
%{\bf 25}, 25 (2002).

%\bibitem{Capelle} K. Capelle, E. K. U. Gross, {\it Phys. Rev. B}
%{\bf 59}, (10) (1999).

%\bibitem{Cabrera} B. Cabrera, H. Gutfreund, W. A. Little, {\it
%Phys. Rev. B} {\bf 25}, (11) 6644 (1982).

%\bibitem{DeWitt} B. S. DeWitt, {\it Phys. Rev. Lett.} {\bf 16},
%1092 (1966).

%\bibitem{Ross} D. K. Ross, {\it Jour. Phys. A} {\bf 16}, 1092
%(1983).

\end{thebibliography}
\end{document}